\documentclass[a4paper]{jpconf}
\usepackage{graphicx}
\usepackage{citesort}
\begin{document}
\title{Progress toward Barium Tagging in High Pressure Xenon Gas with Single Molecule Fluorescence Imaging}

\author{N. Byrnes, F. W. Foss Jr., B.J.P Jones, A.D. McDonald, D.R. Nygren, P. Thapa and A. Trinidad}

\address{University of Texas at Arlington, Arlington, TX 76019}
\ead{ben.jones@uta.edu}

\author{On behalf of the NEXT collaboration.}

\begin{abstract}
We present an update on the development of techniques to adapt Single
Molecule Fluorescent Imaging for the tagging of individual
barium ions in high pressure xenon gas detectors, with the goal of realizing
a background-free neutrinoless double beta decay technology. Previously reported
progress is reviewed, including the recent demonstration
of single barium dication sensitivity using SMFI. We then describe
two important advances: 1) the development of a new class of custom
barium sensing fluorescent dyes, which exhibit a significantly
stronger response to barium than commercial calcium sensing compounds in aqueous
solution; 2) the first demonstration of
a dry-phase chemosensor for barium ions.  This proceeding documents work presented at the 9th Symposium on Large TPCs for Rare Event Detection in Paris, France.
\end{abstract}

\section{Barium Tagging in High Pressure Xenon Gas with SMFI}

A convincing detection of neutrinoless double beta decay ($0\nu\beta\beta$)
would demonstrate that the neutrino is its own antiparticle. Such
a discovery would prove that the standard model is a low energy effective
theory; demonstrate a new mechanism of mass generation that explains
the disparity of charged and neutral lepton mass scales; prove that
lepton number is not a symmetry of nature; and lend weight to leptogenesis
as a viable explanation for the matter-antimatter asymmetry of the
Universe. The search for $0\nu\beta\beta$ is accordingly considered
to be a top scientific priority in nuclear and particle physics worldwide. 

To realize half-life sensitivities nearing $10^{28}$
yr, an ideal experiment would be a fully active, ultra-low background,
ton-scale or larger detector with strong positive signal criteria capable of providing an unambiguous
discovery claim.  For detectors using the isotope $^{136}$Xe, it has long been recognized
that efficient and selective detection of the daughter nucleus $^{136}$Ba,
in coincidence with electron energy measurements of precision better than $\sim$2\%
FWHM to reject the two-neutrino double beta decay background, would represent such a technology.

R\&D on techniques for ``barium tagging'' has been pursued actively for
both liquid and gaseous xenon experiments for at least 17 years~\cite{Moe:1991ik}.  Several single-atom or single-ion-sensitive detection methods
exist, emerging from various disciplines in physics and chemistry.
The question facing the field is: can any of these techniques be implemented
with high efficiency, in a liquid or gaseous xenon environment? 

An important consideration when assessing barium tagging approaches 
concerns the charge state of the daughter nucleus.  Barium from double 
beta decay is born in a highly ionized state Ba$^{N+}$ that quickly captures electrons
from neutral Xe until further capture is energetically disfavored, stopping at Ba$^{++}$.
In liquid xenon, recombination with locally thermalized electrons then produces further neutralization, producing an ensemble of ionic and atomic species including Ba and Ba$^{+}$ \cite{Albert:2015vma}. The lack of recombination in gaseous xenon, on the other hand, implies that Ba$^{++}$ will be the dominant outcome~\cite{Novella:2018ewv}.  Distinct technologies thus appear optimal for these two cases.

Barium tagging approaches for liquid xenon \cite{Sinclair:2011zz,Mong:2014iya} have traditionally focused on cycling fluorescence transitions of the outer electron in Ba$^+$.  Ba$^{++}$, on the other hand, has a noble-like electron configuration without low-lying fluorescence transitions.
To detect Ba$^{++}$ using fluorescence techniques it is necessary
to add such transitions artificially. A method for achieving this was proposed in \cite{Elba} and further developed in~\cite{Jones:2016qiq}, using the technique of Single Molecule Fluorescence Imaging (SMFI). This method is widely used in
biochemical microscopy for sensing Ca$^{++}$ ions in biological media \cite{stuurman2006imaging,fish2009total,Thomas2000,Lu2007,Oliver2000,nakahara2005new}.
In SMFI, a specially designed molecule is employed that is non-fluorescent
in isolation but becomes fluorescent upon chelation of a target
ion. The fluorescence enabled upon binding can be observed by probing
with an excitation laser and collecting the longer wavelength fluorescence emission using highly sensitive EM-CCD cameras.

\begin{figure}[t]
\begin{centering}
\includegraphics[width=0.9\columnwidth]{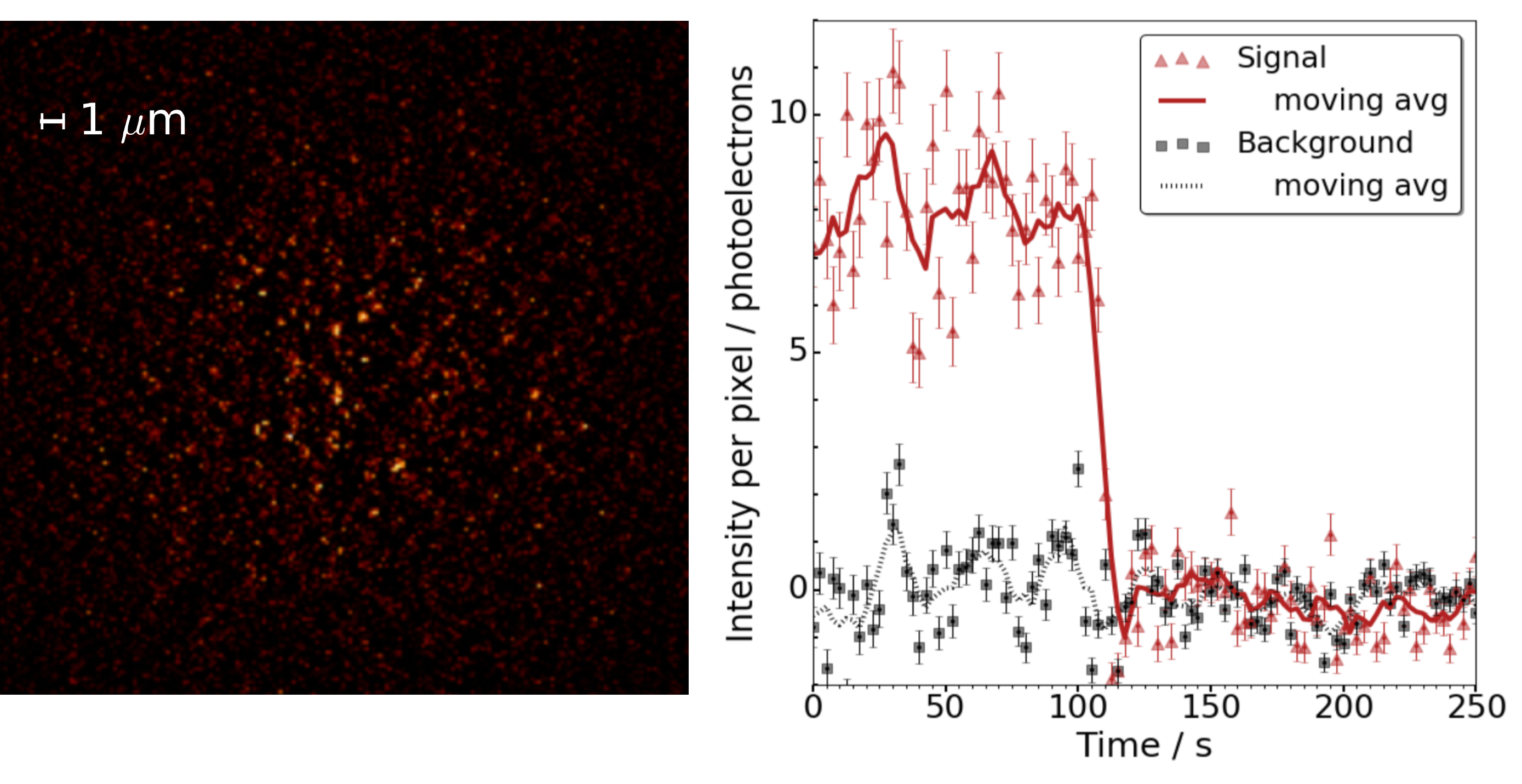}
\par\end{centering}
\caption{Left: TIRFM image showing individual Ba$^{++}$ ions near the sensor surface; Right: The time profile of a single spot, showing the discrete photo-bleaching process that is
the hallmark of single molecule fluorescence. Reproduced
from~\cite{McDonald:2017izm}.\label{fig:PRLPlots}}
\end{figure}

The NEXT collaboration has been developing SMFI-based
barium tagging since 2015. In early exploratory work, we demonstrated
that commercially available molecules FLUO-3 and FLUO-4, designed
for Ca$^{++}$ detection, are sensitive probes for Ba$^{++}$, thus making them promising barium tagging agents~\cite{Jones:2016qiq}. Using
total internal fluorescence microscopy~\cite{axelrod2003} (TIRFM) with FLUO-3, the first single
barium dication fluorescence detection was then demonstrated~\cite{McDonald:2017izm}. Individual ions were spatially resolved with 2 nm super-resolution and identified
with 13~$\sigma$ statistical significance over background via their
sharp photo-bleaching transitions (Fig.~\ref{fig:PRLPlots}). 

We have also studied the drift and survival properties
of the dication state in high pressure Xe gas
theoretically~\cite{Bainglass:2018odn} and this is the subject of an ongoing experimental
program. Also ongoing are the development of ion concentration methodologies
at the cathode and creation of dry-compatible fluorophores for ion
imaging in Xe gas.

\begin{figure}
\begin{centering}
\includegraphics[width=0.9\columnwidth]{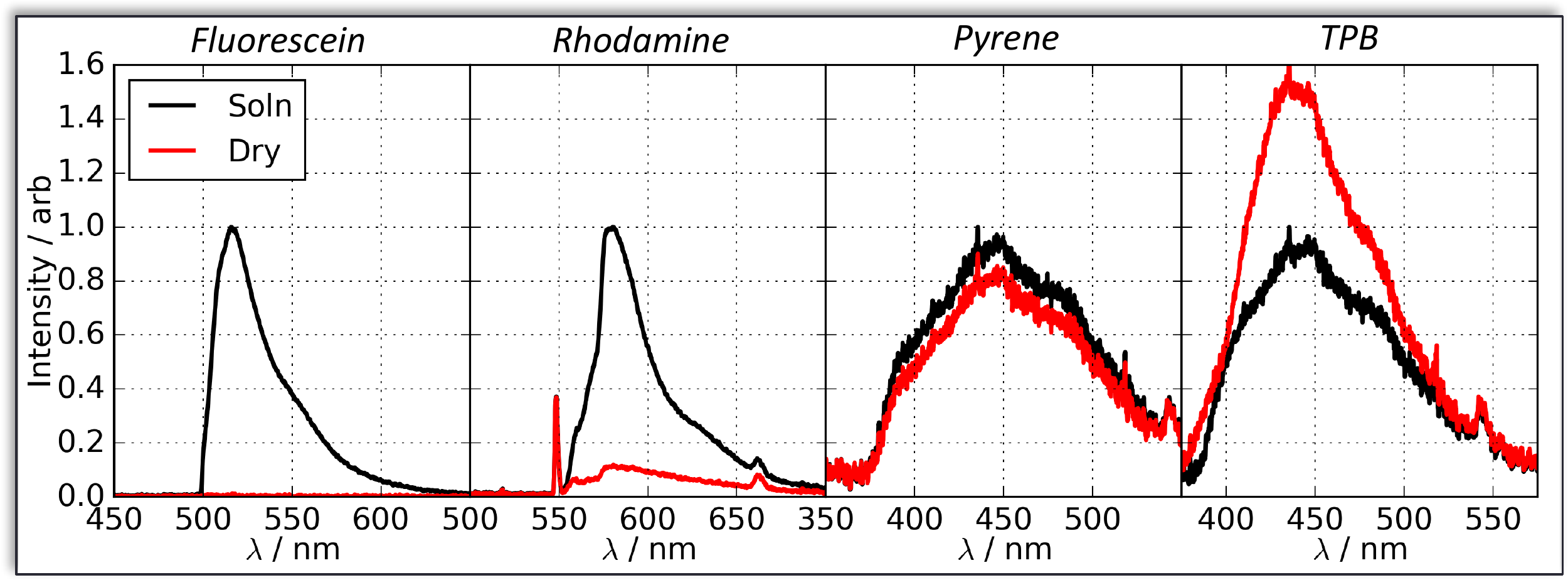}
\par\end{centering}
\caption{Wet and dry response of various fluors that may be used in SMFI molecules. \label{fig:WetDryPlots}}
\end{figure}

\section{Requirements for a Dry-Phase SMFI Agent}

\begin{figure}
\begin{centering}
\includegraphics[width=0.99\columnwidth]{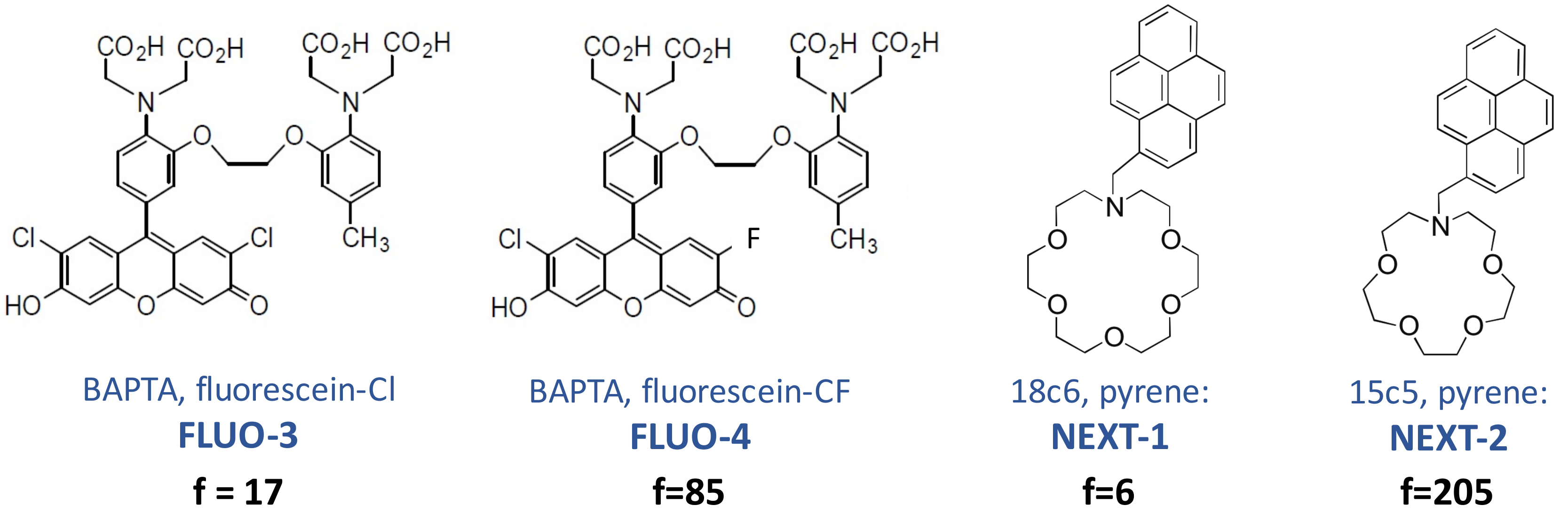}
\par\end{centering}
\caption{Molecules described in this proceeding. FLUO-3 and
FLUO-4 are commercially available calcium sensors and NEXT-1 and NEXT-2
are our own custom-made molecules.

  \label{fig:Molecules}}
\end{figure}

Fluorophores FLUO-3 and FLUO-4 comprise of a BAPTA-like receptor
bonded onto a fluorescein-like fluor, shown in Fig \ref{fig:Molecules}, left. The on-off fluorescence response in the FLUO family is understood to arise
from quenching of the fluorescein group, presumably by photoinduced electron transfer from
the lone pair of electrons on nitrogen. When unbound to an
ion, these electrons may move freely into the fluorescein group to
inhibit fluorescence. In the presence of a Ba$^{++}$, however, the
BAPTA-like receptor binds the ion in a cage-like structure, with
its electrostatic force pulling the nitrogen electrons into a bonding-type
configuration described as Lewis acid-base complexation. This inhibits the movement of nitrogen's lone pair of electrons into the fluorescein group, and prevents the flourescent response from being quenched.

The FLUO family of chemosensors is typically deployed in aqueous solutions. For example, our previous work~\cite{McDonald:2017izm} used
FLUO-3 suspended in a PVA matrix with liquid pockets to resolve Ba$^{++}$
ions from barium perchlorate solution. Two unfortunate features
appear to render this family of molecules inappropriate for use in
a dry environment: 1) the binding to Ba$^{++}$ requires deprotonation
of the four carboxylic acid groups in the BAPTA-like receptor, which is not expected
to occur in a dry state. 2) While fluorescein is a bright fluor in solution,
 our data (Fig.~\ref{fig:WetDryPlots}, first panel) show that fluorescence is suppressed when dried. 

To address these issues, we have designed and synthesized custom SMFI molecules based off known alkaline earth metal binding crown ethers that avoid these problems.   First, we have identified binding groups based on aza-crown ethers and aza-cryptands as good candidates for dry Ba$^{++}$ capture, since they can bind the ion strongly but without deprotonation; Second, other
fluors such as pyrene exhibit less sensitivity to the solvent and microenvironment  (Fig.~\ref{fig:WetDryPlots}, third panel).
 Our hypothesis is that connection of these two building blocks via a nitrogen ``switch'' similar to that used in FLUO-3 and FLUO-4 should yield molecules with a strong on-off response to Ba$^{++}$ in dry environments.

\section{Preliminary Results with custom-made aza-crown derivatives}

Based on the considerations above, we have embarked upon systematic
study of aza-crown and aza-cryptand-based binders coupled to a variety
of fluorophores, starting with pyrene and subsequently investigating similarly promising fluorophores naphthalimides,  boron-dipyrromethenes and anthracene. Here we present early results using mono-aza-crown ethers (15-crown-5 and 18-crown-6) linked to methylpyrenes, 15-crown-5-pyrene and 18-crown-6-pyrene.  Also under development are 21-crown-7 receptors, though we do not present results from those here.

Synthesis of 15-crown-5 ether (15c5) and 18-crown-6 ether (18c6) binding domains is achieved in parallel, via ether synthesis under phase transfer conditions \cite{luk2012synthesis} with N-benzyl-bisethanolamine acting as the linchpin that reacts at two ends of various length polyethers. Hydrogenolysis of the N-benzyl protecting group yields unmodified aza-crown ethers ready for the installation of various fluorophores. The late installation of fluorophores via S$_N$2 substitution allows for a divergent synthetic pathway to a wide variety of potentially successful fluorescent barium tags without significantly altering the synthesis \cite{nakahara2005new}.

\begin{figure}
\begin{centering}
\includegraphics[width=0.65\columnwidth]{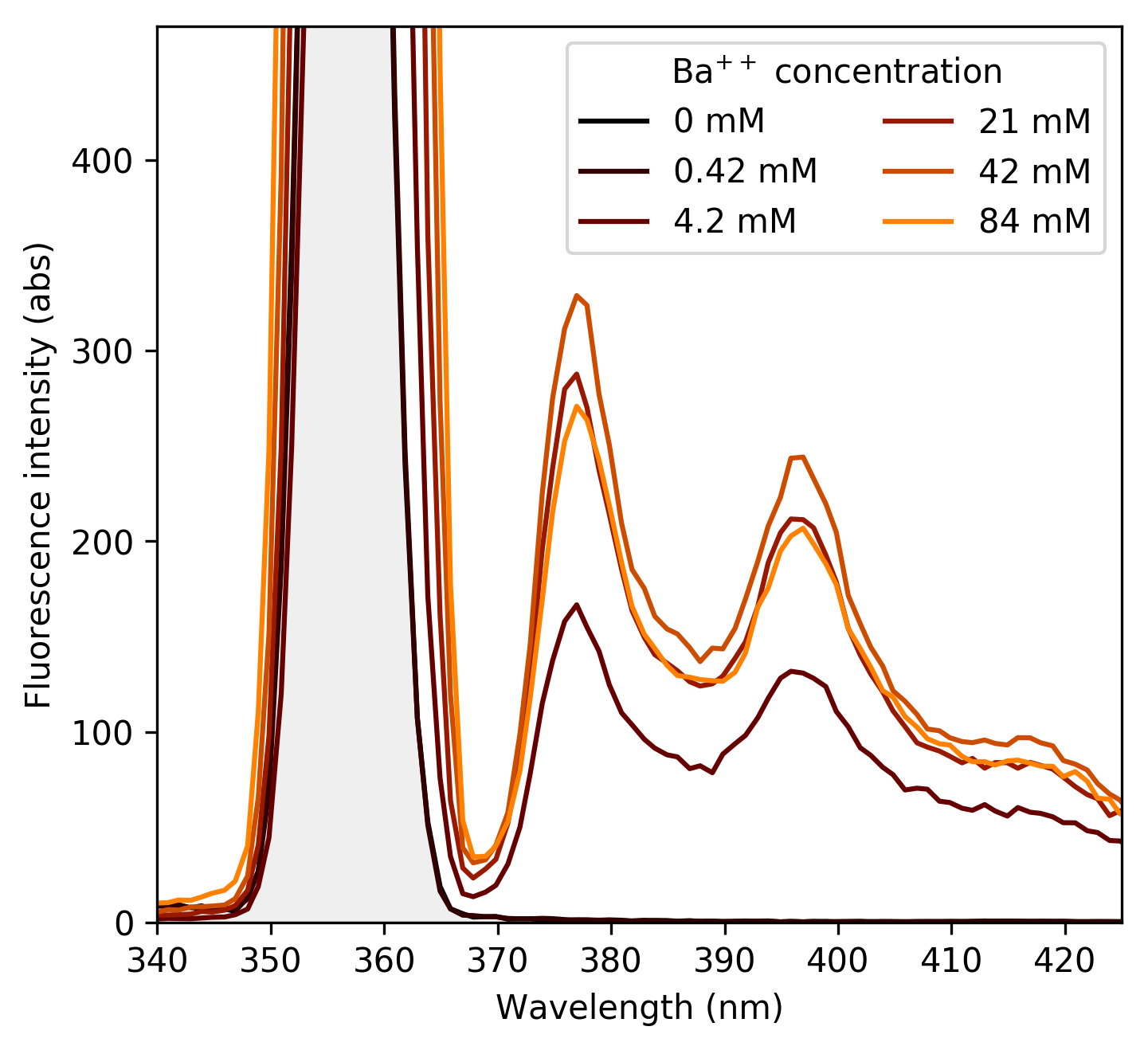}
\par\end{centering}
\caption{Response of NEXT-2 dye with steadily increasing Ba$^{++}$
concentrations.  The light grey shaded area represents the response in the ``off'' state, primarily from scattered excitation light.   \label{fig:Fluorescence-response-to}}
\end{figure}

Fig. \ref{fig:Fluorescence-response-to} shows the Ba$^{++}$ response
of the most promising of our molecules to date, based on 15c5 and
pyrene, which we have named NEXT-2 and show in Fig.~\ref{fig:Molecules}, right. Initially, buffered aqueous solutions of fluorophores were prepared by serial dilution and fluorescent response
was measured using a Cary Agilent Eclipse fluorescence spectrometer.
The molecule is excited at its lowest energy $\lambda_{max}$ (357 nm), near the absorption peak of pyrene, and emits a broad, two-peaked spectrum at 375 and 400 nm when bound to Ba$^{++}$.

The performance of SMFI agents is often characterized by fluorescence
ratio, $f=\left(F_{max}-F_{min}\right)/F_{min}$, where $F_{max}$
is the fluorescence intensity when saturated with ions and $F_{min}$
is the minimum observable intensity when uncontaminated. Integrating
the spectral region above 375 nm, NEXT-2 exhibit a fluorescence ratio
exceeding 200 for barium. This is to be compared to the commercially
available dyes FLUO-3 and FLUO-4, where our previous studies found
$f=17$ and $f=85$, respectively. Remarkably, this custom-made molecule exhibits a
much stronger transition between the on- and off-states than the
the commercial molecules we tested for Ba$^{++}$ sensitivity in~\cite{Jones:2016qiq}. The closely related molecule based
on 18c6 and pyrene (which we have named NEXT-1, and show in Fig.~\ref{fig:Molecules}, second from right), also exhibits barium
sensitivity although less strongly, with $f=6$. 

The demonstration of strong barium response from custom made SMFI
agents in our laboratory is an important step in this R\&D program,
opening a path to systematic exploration of a wide variety of related molecular
structures for SMFI-based barium tagging in xenon gas. 

\section{Preliminary Dry Fluorescence Response}

The primary goal of custom SMFI fluorophore development is not brightness
optimization, but performance in a dry environment. No
commercially available calcium sensors that we have tested have satisfied
this requirement. To assess the fluorescence of our custom molecules
in a dry environment, we used the microscopy setup described in~\cite{McDonald:2017izm}
with dried samples of fluorophore, with and without added Ba$^{++}$.
Due to the non-availability of ultraviolet wavelength dichroic mirrors
and filters, excitation is delivered at 400~nm with a dichroic separation
of excitation and emission at 409~nm. This is far from ideal for these
molecules, which would respond much more strongly with a shorter wavelength
excitation and fluorescence collection integrating above 375~nm. Nevertheless,
we established that with sufficiently long microscope exposures, fluorescence
from raw pyrene can be observed in this configuration, both in bulk
solution and at the single molecule level. This demonstrates that
even with significantly off-peak excitation, the system is sufficiently
sensitive for these tests.

\begin{figure}
\begin{centering}
\includegraphics[width=0.9\columnwidth]{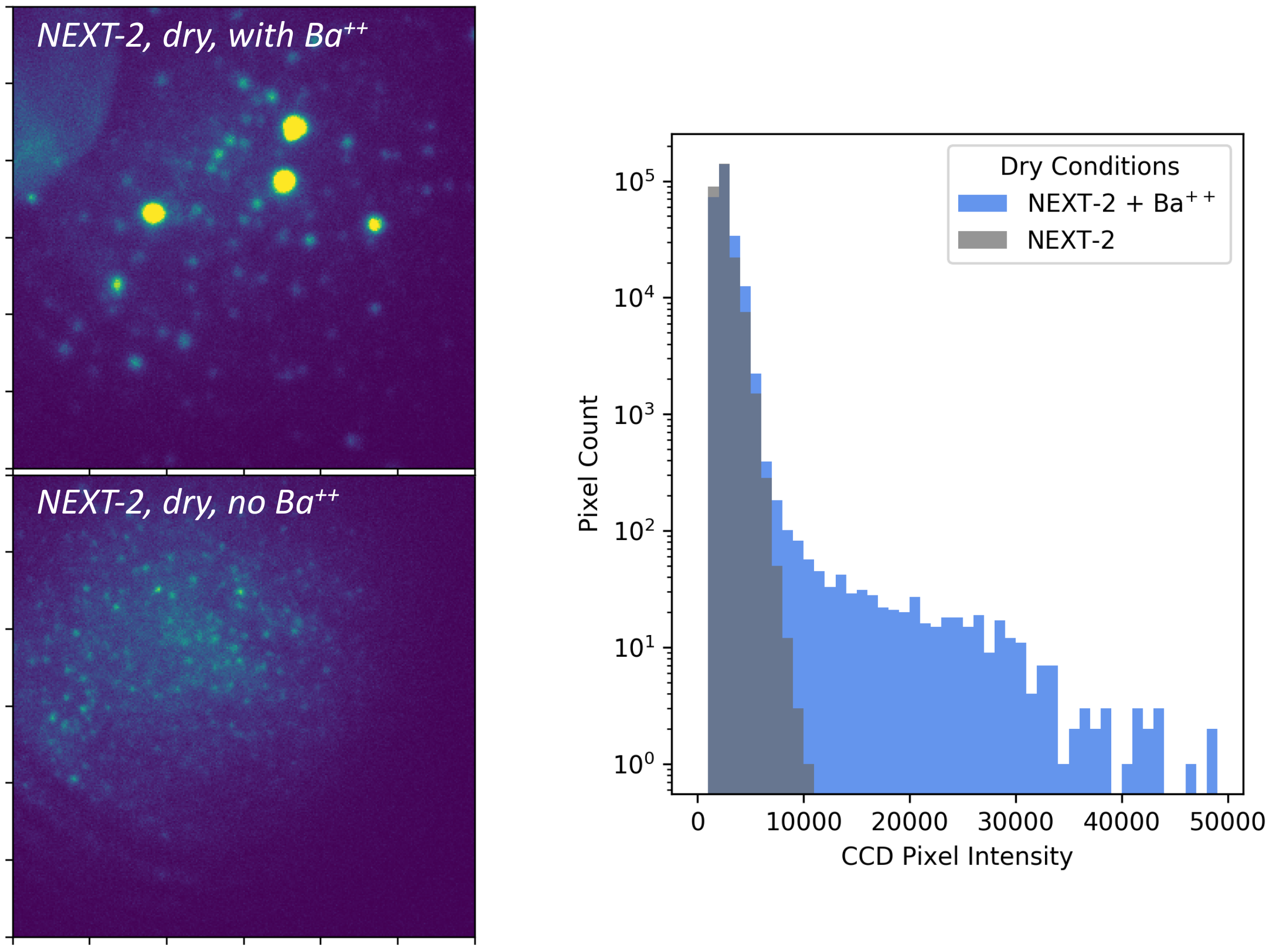}
\par\end{centering}
\caption{First results from dry fluorescence studies of NEXT-2 fluorophore.
Left: microscope images of barium-spiked and barium un-spiked samples;
Right: CCD pixel intensity histogram for these two cases. \label{fig:DryFig}}

\end{figure}

To test for dry fluorescence, samples of barium-spiked and barium-un-spiked
NEXT-1 and NEXT-2 dyes were dried onto microscope slides. These slides
were then scanned for fluorescence, with emission collected by Hamamatsu ImagEM X2 EM-CCD camera. Example images are shown in Fig.~\ref{fig:DryFig}, left, with the corresponding pixel intensity distributions shown in Fig.~\ref{fig:DryFig}, right. 

The un-spiked samples are optically quiet, showing a background of
small isolated spots of low-level fluorescence on a smooth background
of scattered light. Barium-spiked samples, on the other hand, exhibit
very bright fluorescent regions. These are understood to be macroscopic
clusters of fluorescent material rather than single molecules, based
on their lack of photo-bleaching and high fluorescence intensity.  Similar behaviour was observed for both NEXT-1 and NEXT-2.  A full account of these data and their analyses will be forthcoming in a future publication.

These data demonstrate a strong fluorescent response in the presence of chelated
Ba$^{++}$ ions in the dry phase.  Single molecule detection was not achieved in this test because of 
the imperfect preparation of the SMFI layer, leading to crystals.
In the next stage of this program we will deposit modified fluorophores in a
well-controlled monolayer using molecular self-assembly via siloxane surface tethers.
This will allow individual fluorescent molecules to be resolved, rather than large clusters.

\section{Conclusion and Outlook}

In this talk and proceeding we have summarized previous work on barium tagging 
using SMFI for the NEXT experiment.  We presented two important new advances: 1) the first
demonstration of fluorescence response from custom-made barium sensing
molecules from our laboratories that exceeds the barium-induced response
from commercial calcium sensors; 2) the first demonstration of a barium
chemosensor that maintains responsiveness in the dry phase. A fully
systematic exploration of the fluorescent response of 15c5, 18c6 and
21c7 derivatives with a variety of quenchable fluors in wet and dry
environments is presently under way.

In addition, work is commencing to produce a well-controlled monolayer
via self-assembly of SMFI molecules onto microscope slides, with the
goal of producing a high quality, robust, uniform, single-ion-resolving sensing
surface. This builds upon our previous demonstration of single ion
identification in aqueous suspensions, translating the approach to
thin layer of immobilized molecules on a sensitized glass or quartz
surface. We will proceed to test the response of this layer under bombardment
of barium ions in a gaseous xenon environment, to demonstrate their
viability as SMFI agents within the unfamiliar environment of a xenon
gas detector.

Following successful demonstration of single barium dication resolution
with SMFI and creation of a dry-compatible barium chemosensor, the
identification of barium ions in high pressure xenon gas may be close
at hand. If realized, this could yield a new technology to enable large,
effectively background-free $0\nu\beta\beta$ experiments, that would
represent a powerful tool for discovery of this elusive process.

\section*{Acknowledgements}

We thank Katherine Woodruff and Fernando Cossio for comments on this manuscript.  The work described in this proceeding was supported by the Department of Energy under Early Career Award number DE-SC0019054. The University of Texas at Arlington group is also supported by Department of Energy Award DE-SC0019223.  The NEXT Collaboration acknowledges support from the following agencies and institutions: the European Research Council (ERC) under the Advanced Grant 339787-NEXT; the Ministerio de Econom\'ia y Competitividad of Spain under grants FIS2014-53371-C04 and the Severo Ochoa Program SEV-2014-0398; the GVA of Spain under grant PROMETEO/2016/120; the Portuguese FCT and FEDER through the program COMPETE, project PTDC/FIS/103860/2008; the U.S. Department of Energy under contracts number DE-AC02-07CH11359 (Fermi National Accelerator Laboratory) and DE-FG02-13ER42020 (Texas A\&M) and DE-AC02-06CH11357 (Argonne National Laboratory).

\section*{References}
\bibliography{biblio}

\end{document}